\documentclass{PoS}
\newcommand{\MSb}{$\overline{\rm MS}$ }

\newcommand{\gv}{\mbox{GeV}}

\title{About the EW contribution to the relation between pole and
$\overline{\mbox{MS}}$ masses of the top-quark 
in the Standard Model}

\ShortTitle{EW contribution to the running  mass of top-quark}

\author{Fred Jegerlehner\\
{\normalsize Humboldt-Universit\"at zu Berlin, Institut f\"ur Physik,} \\
{\normalsize  Newtonstrasse 15, D-12489 Berlin, Germany}\\
{\normalsize Deutsches Elektronen-Synchrotron (DESY), Platanenallee 6, D-15738 Zeuthen, Germany} }

\author{\speaker{Mikhail Kalmykov}\thanks{This work was supported in part 
by the German Federal Ministry for Education and Research BMBF through Grant No.\ 05~H12GUE, 
by the German Research Foundation DFG through the Collaborative Research Center No.~676 
{\it Particles, Strings and the Early Universe---The Structure of Matter and Space-Time} 
}\\
{\normalsize II. Institut f\"ur Theoretische Physik, Universit\"at Hamburg,}\\
{\normalsize Luruper Chaussee 149, D-22761 Hamburg, Germany} \\
        E-mail: \email{kalmykov.mikhail@gmail.com}}

\author{Bernd~A.~Kniehl \\
{\normalsize II. Institut f\"ur Theoretische Physik, Universit\"at Hamburg,}\\
{\normalsize Luruper Chaussee 149, D-22761 Hamburg, Germany} }

\abstract{
Results of our recent re-analysis of the electroweak contribution to
the relation between pole and running masses of top-quark within
the Standard Model is reviewed.  We argue, that if vacuum of SM is stable,
then there exists an optimal value of renormalization group scale
(IR-point), at which the radiative corrections to the matching
condition between parameters of Higgs sector and pole masses is
minimal or equal to zero. Within the available accuracy, we find the
IR-point to lie in an interval between value of Z-boson mass and twice
the value of W-boson mass. The value of scale is relevant for
extraction of Higgs self-coupling from cross-section as well as for
construction of effective Lagrangian. }

\FullConference{XXI International Workshop on Deep-Inelastic Scattering and Related Subject -DIS2013,\\
		22-26 April 2013\\
		Marseilles,France}

\begin{document}

\noindent
{\bf Motivation.}  After the discovery of the Higgs boson~\cite{Higgs}
- the last important building block of the Standard Model (SM)
required by its renormalizability \cite{SM}- and the still missing
direct detection of new physics beyond SM at the LHC, the
self-consistency of the SM has attracted a lot of notice.  The key
question is the stability of the Higgs potential when extending SM
physics to such high scales as the Planck mass, where we know gravity
must come into play and SM physics alone cannot lead further.  It
would be interesting to determine the scale at which SM may break
down~\cite{ellis}. One of the approaches to answer this question is
based on the renormalization group (RG) analysis of the SM running
couplings, specifically of the Higgs self-coupling $\lambda$ and the
question whether it stays positive up to the Planck scale which would
imply the vacuum to remain stable. The recent re-analysis of the stability
of vacuum performed in~\cite{RGE-analysis-1,RGE-analysis-2}, with 3-loop RG functions evaluated in
\cite{RG1,RG2,RG3}, revealed a surprisingly small ``critical value''
of Higgs boson mass (i.e. the stability bound), $M_H^{\mbox{cr.}} \sim
129~\mbox{GeV}$. The main difference between~\cite{RGE-analysis-1}
and~\cite{RGE-analysis-2} was related to the uncertainties adopted for
the input parameters $\delta M_H^{\mbox{cr.}}  = \pm 6~\mbox{GeV}$ in
\cite{RGE-analysis-1} versus $\delta M_H^{\mbox{cr.}}  = \pm
1.5~\mbox{GeV}$ in~\cite{RGE-analysis-2}. These stability bounds
sensitively depend on the value of pole mass of top-quark, $M_t$
and/or its \MSb version $m_t$ and therefore a careful evaluation of
the relationship between $M_t$ and $m_t$ is mandatory.  An updated
analysis of the determination of $M_t$ and its uncertainties has been
presented in~\cite{ADM-analysis}. It has been pointed out, that $M_t = 173.1
\pm 0.7~\mbox{GeV}$, which has been used in~\cite{RGE-analysis-2} as
an input, does not relate directly to the value of pole mass
of the top-quark (see also the discussion in~\cite{Hoang}). Adopting the updated
value~\cite{ADM-analysis}, $M_t =
173.3 \pm 2.8~\mbox{GeV}$, (an indirect determination
yielded 
$M_t=175.7^{+3.0}_{-2.2}~\mbox{GeV}$ ~\cite{karlsruhe}
or 
$M_t=175.8^{+2.7}_{-2.4}~\mbox{GeV}$ ~\cite{gfitter}
) the results
of~\cite{RGE-analysis-1} and~\cite{RGE-analysis-2} are close to each
other: $\delta M_H^{\mbox{cr.}}  = \pm 5.6~\mbox{GeV}$ (see also the
detailed discussion of~\cite{RGE-analysis-1,
RGE-analysis-2,ADM-analysis} in~\cite{Masina}).  In the case of
stability of vacuum~\cite{RGE-analysis-1,RGE-analysis-2,JKK,Fred1}, the SM
works up to the Planck scale and new physics is not necessary to cure
for the instability of the vacuum~\footnote{The new analysis is presented in \cite{strumia13}}. 
The interrelation between the SM
and gravity/cosmology have been discussed in~\cite{BS,SW}.

The value of the pole-mass of top-quark has been extracted from
measurements of the hadronic cross section $\sigma_{pp \to t\bar{t}}$
at the Tevatron and the LHC: for example, $M_t = 173.18 \pm
0.94~\mbox{GeV}$ \cite{Tevatron}, $M_t = 173.3 \pm 1.4~\mbox{GeV}$
\cite{deliot}, $M_t = 172.31 \pm 1.55~\mbox{GeV}$ \cite{ATLAS-top},
$M_t=176.7^{+3.8}_{-3.4}~\mbox{GeV}$ \cite{CMS}.
Recent theoretical QCD predictions, like the results of the direct
perturbative evaluation of QCD NNLO corrections to the total inclusive
top-quark pair production cross-section at hadronic
colliders~\cite{CM}, can be improved by inclusion of the electroweak
(EW) corrections~\cite{Beenakker1993}.  Numerically, the EW
contribution to the hadronic $t\bar{t}$ production is of the order of
a few percent. Differential cross sections can be affected
substantially by EW corrections especially with increasing LHC
energies~\cite{EW-review}.

The standard parametrization of cross-sections is based on the
on-shell scheme with pole-masses of particles as input
parameters. However, the pole mass of a quark suffers from renormalon
contributions~\cite{renormalon}, which give rise to a slow convergence
of the perturbative expansion for any physical observable parametrized
in terms of on-shell parameters. Moreover, due to the confinement of
quark and gluons, the pole mass of a quark does not have an evident
physical meaning. Early in the history of perturbative QCD
calculations it has been noticed that the \MSb parametrization has
better convergence properties and in addition is much simpler (mass
independent) in describing the scale dependence governed by the
renormalization group equations. The advantage of utilizing the
$\overline{\rm MS}$ mass (running mass) $m_t$ of the top-quark as input
parameter for parametrization of top-pair production has been explored
in~\cite{LMU2009}. It has been shown in by direct calculations, that
there is an essential reduction of the scale dependence as well as a
faster convergence of the perturbative expansion for the $t\bar{t}$
production cross section. If fact, the NLO and NNLO corrections are
much smaller in the \MSb parametrization. The same properties
- stability of scale dependence and fast convergence of perturbative
series - are valid also for differential distributions~\cite{DM}. To
include the EW contribution~\cite{Beenakker1993} in a corresponding
way as performed in~\cite{LMU2009}, the EW corrections in the relation
between pole and \MSb masses should be also included.

\noindent
{\bf The EW contribution to the running mass of the top-quark.}  For
the low energy experiments at LEP energies, the splitting of full SM
corrections into QED-, weak- and QCD-contributions has been quite
reasonable - within NLO accuracy, the mixing effects do not play an
essential role.  Moreover, the QED and QCD corrections are
``universal'', in the sense that they depend mainly on the number of
fermions and on their masses. However, in order to achieve percent
level precision theoretical predictions not only QCD NNLO radiative
corrections should be included. The EW part as well as mixing EW
$\times$ QCD corrections have to be included in a systematic way.  For
example, the QCD interaction is not responsible for the non-zero width
of top-quark, which can be understood precisely only by inclusion of
EW interaction. In any case, this non-zero width (EW effect) should be
included in QCD corrections~\cite{CM,LMU2009,Beenakker1993} (which has
not yet been done) and can modify the prediction up to $1 \%$ (the
technique of~\cite{passarino} can be directly applied in this case).
%
%

In~\cite{JKK} we have evaluated the EW contribution to the relation
between pole- and running-mass of the top quark within the SM. The main effect is due
to the matching conditions between $M_t$ and $m_t$ at the
scale $M_t$. In addition, the running of parameters has to be taken into
account. The quark mass anomalous dimension $
\mu^2 \frac{d}{d \mu^2} \ln m_q^2(\mu^2) = \gamma_q(\alpha_s,\alpha)\;,
$
has two parts: the QCD and the EW contribution,
$
\gamma_q(\alpha_s,\alpha)
=
\gamma_q^{\rm QCD}
+
\gamma^{\rm EW}_q \;,
$ where $\gamma_q^{\rm QCD}$ includes all terms which are proportional
to powers of $\alpha_s$ only and $\gamma^{\rm EW}_q$ includes all
other terms proportional to at least one power of $\alpha$, and beyond
one loop multiplied by further powers of $\alpha$ and/or
$\alpha_s$. It has been shown in Ref.~\cite{JKV}, that $\gamma^{\rm
EW}_t$ in the $\overline{\rm MS}$ scheme may be written in terms of RG
functions of parameters in the unbroken phase of the
SM~\cite{RG1,RG3}, $
\gamma^{\rm EW}_t = \gamma_{y_t} + \frac{1}{2} \gamma_{m^2} - 
\frac{1}{2} \frac{\beta_\lambda}{\lambda} \;,
$ where $y_t$ is the Yukawa coupling of the top-quark, $\gamma_{y_t}
\equiv \mu^2 \frac{d}{d \mu^2} \ln y_t$, and $m^2$ and $\lambda$ are
the parameters of the scalar potential\footnote{The bridge between the
UV mass counterterms of the masses $m_H^2,m_t^2,m_W^2$ evaluated
in~\cite{JKV,JK,RGE-analysis-1} in the broken phase of the SM and the
RG equations for the SM parameters $m^2,\lambda,y_t^2$ in the unbroken
phase~\cite{DRTJones}, gives rise to the following parametrization, in
terms of the effective Higgs vacuum expectation value $v(\mu^2)$:\\
$\sqrt{2}G_F = 1/v^2 = 1/(246.22)^2~\gv^{-2}, \quad 1/v^2(\mu^2) =
\sqrt{2} G_F^{\overline{MS}}(\mu^2)$, \\ $m_H^2(\mu^2) = 2 m^2 (\mu^2)
\;, \quad
\lambda(\mu^2) = \frac{3 m_H^2(\mu^2)}{v^2(\mu^2)} \;, \quad   
y_t^2(\mu^2) = \frac{2 m_t^2(\mu^2)}{v^2(\mu^2)} \;, \quad g^2(\mu^2)
= \frac{4 m_W^2(\mu^2)}{v^2(\mu^2)}$.} $ V = \frac{m^2}{2} \phi^2 +
\frac{\lambda}{24} \phi^4.  $ The values of $\gamma_{y_t}$ and
$\gamma_{m^2}$ are bounded under scale variations $M_W < \mu <
M_{\mbox{Planck}}$~\cite{JKK,strumia,Fred1}, $\beta_\lambda$ is
negative, such that $\lambda$ is decreasing with increasing value of
$\mu$. According to estimates in~\cite{JKK, Fred1} $\beta_\lambda$ has a zero
somewhat below the Planck scale while $\lambda$ is still positive and
then slightly increases before $\mu$ reaches
$M_{\mbox{Planck}}$.
 
\noindent
{\bf The evaluation of the matching conditions.} Since masses are
generated by the Higgs mechanism, the decoupling theorem~\cite{AC} is
not valid in the week sector of the SM, such that a top quark-mass
dependence cannot be renormalized away even at such low scales as $\mu
\sim M_W$. As a consequence, the complete set of particles and the
corresponding diagrams must be evaluated, especially for the
top-quark.  The numerical value of the EW contribution to the relation
between pole and \MSb mass of a quark may be extracted from the \MSb
renormalized propagator~\cite{FJ81} and have been evaluated
analytically\footnote{Two terms on the second line of Eq.~(12)
in~\cite{JK-erratum} are to be modified:\\ "$-\frac{m_t^4}{m_H^4} \ln(1
+ y) +
\frac{m_t^2}{m_H^2} \frac{3 + y^2}{1 + y} \ln y$" should read
"$-\frac{m_H^4}{m_t^4} \ln(1 + y) +
\frac{m_H^2}{m_t^2} \frac{3 + y^2}{1 + y} \ln y$".} in~\cite{HK,JK}. 
For the finite part of the $O(\alpha \alpha_s)$ mixing-correction,
numerical agreement with the semi-analytic result of~\cite{S2005} has
been established. Intermediate results of~\cite{JK} have been
cross-checked in~\cite{KK}. The complete $O(\alpha^2)$ EW contribution
is not yet available, and results presented in~\cite{kuhn}
and~\cite{martin} are not in agreement. Using an indirect estimation
of the $O(\alpha^2)$ corrections, we evaluated in~\cite{JKK} that for
$124 < M_H < 126 ~\mbox{GeV}$ the EW contribution is large and has
opposite sign relative to the QCD contributions, so that the total SM
correction is small and approximately equal to $m_t(M_t)-M_t \sim 1
\pm O(1) ~\gv.$\footnote{The results of~\cite{kataev-kim} are not
relevant for our analysis.}

\noindent
{\bf An optimal value for the RG matching scale.}  At the Planck scale
perturbative SM RG-evolution stops to make sense, since in any case
gravity contributions would come into play~\cite{SW}. When considering
matching conditions for a fixed pole mass $M$ versus a running \MSb
mass $M(\mu^2)$ it is evident that at some scale, we will call it ``IR
scale'', we must have $m(\mu_{\rm IR}^2)=M$. Depending of the sign of
the correction $m(M)-M$ and the sign of $\mu^2\, d m(\mu^2)/d \mu^2$
at scale $M$ one finds $\mu_{\rm IR}<M$ or $\mu_{\rm IR}>M$. An
interesting question then is whether there exists a preferable
$\mu_{\rm IR}$ and how to define it? An optimal value for $\mu_{\rm
IR}$ could be defined as value of $\mu$ where the radiative
corrections to the some (or all) matching conditions between running
coupling and pole masses are minimal, close or equal to zero. Since
the main interest of our RG-analysis concerns the Higgs sector of the
SM, we will analyze the matching conditions for $\lambda(\mu^2)$,
$G_F(\mu^2)$ and $m_H(\mu^2)$. At the one-loop level, the relation
between the running and the physical Fermi constant is (see Eq.(A.3)
in~\cite{RGE-analysis-1}): $$
\frac{G_F(\mu^2)}{G_F} - 1 
 =   
   \frac{g^2(\mu^2)}{16 \pi^2}  f^{(1)}_{G,\alpha}(\mu^2) 
= 
\frac{g^2(\mu^2)}{16 \pi^2}  
\left[
\gamma_{G_F,\alpha} L \!-\! \Delta X_{G_F,\alpha}^{(1)}(\mu^2, M_t,M_H)
\right]
\;,
$$
where $L = \ln \frac{\mu^2}{M_X^2}$ and 
$
\gamma_{G_F,\alpha} 
$
is defined\footnote{In~\cite {JKK}, the r.h.s. of Eq.~(8) 
           should be multiply by the overall factor ``$2 \times G_F^{\overline{\mbox{MS}}}$".}
in~\cite{JKV} as 
$
\gamma_{G_F} 
\!=\! \mu^2 \frac{d}{d \mu^2} \ln G_F(\mu^2) 
\!=\! \frac{\beta_\lambda}{\lambda} \!-\! \gamma_m^2
\!=\! 
\frac{g^2}{16 \pi^2}  \gamma_{G_F,\alpha} \!+\! \cdots.
$
From the condition $f^{(1)}_{G,\alpha}(\mu^2_{\mbox{\rm{IR}}})=0$, we obtain
$$
\frac{\mu_{\mbox{\rm{IR}}}^2}{M_X^2} 
=  \exp\left\{ \frac{\Delta X_{G_F,\alpha}^{(1)}(\mu^2,M_t,M_H)}{\gamma_{G_F,\alpha}} \right\} 
=  \exp\left\{ - \frac{f^{(1)}_{G,\alpha}(\mu^2=M_X^2)}{\gamma_{G_F,\alpha}} \right\} \;.
$$
For $M_t \sim 173-173.5~\gv$ and $124~\gv  \leq M_H \leq 126~\gv$, 
we then find,
$
\mu_{\mbox{\rm{IR}}}^2
\sim  M_t^2 \times 0.3425  \Rightarrow 
\mu_{\mbox{\rm{IR}}} \sim 101.5 ~\mbox{GeV} \;.
$ 
The inclusion of the $O(\alpha \alpha_s)$ correction moves the value
of the IR scale to $\mu_{\mbox{\rm{IR}},G_F}^2$:
\begin{equation}
\mu_{\mbox{\rm{IR}},G_F}^2
\sim   M_t^2 \left[ 0.3075 \!-\! 0.0005 \left( \frac{M_H}{\gv} \!-\! 125 \right)  \right]   
\Rightarrow   
\mu_{\mbox{\rm{IR}},G_F}
= \left[ 96.2 \!-\! 0.01 \left( \frac{M_H}{\gv} \!-\! 125 \right) \pm 5.\ 3 \right]~\gv \;, 
\label{scaleGF}
\end{equation}
which for $M_t=173.5$ and $124~\gv \leq M_H \leq 126~\gv$ and
theoretical uncertainties $\pm 5.3~\gv$, defines the central value of
the IR scale, extracted with $O(\alpha)/O(\alpha \alpha_s)$ accuracy.
At scale $\mu_{\mbox{\rm{IR}},G_F}$, which is close to value of the
Z-boson mass, the running v.e.v. is: $
v\left(\mu_{\mbox{\rm{IR}},G_F}^2\right) = v \equiv 246.22~\gv $ and
the value of the Yukawa coupling $y_q(\mu_{\mbox{\rm{IR}},G_F}^2)$ is
proportional to value of the running mass of quark evaluated at this
scale: $ y_q(\mu_{\mbox{\rm{IR}},G_F}^2) = \frac{\sqrt{2}}{246.22}
\frac{m_q(\mu_{\mbox{\rm{IR}},G_F}^2)}{\gv} $.  Another relation valid
at this scale is:
\begin{equation}
\lambda(\mu_{\mbox{\rm{IR}},G_F}^2) 
= 3 \left( \frac{m_H(\mu_{\mbox{\rm{IR}},G_F}^2)}{ 246.22~\gv} \right)^2\;,
\label{L-at-IR}
\end{equation}
where $ m_H^2(\mu^2) $ is the \MSb mass of the Higgs propagator.

On the same ground we defined the IR scale (we denote it as
$\mu_{\mbox{\rm{IR}},m_H^2}$) from the ratio between the running mass
$m^2$ and the pole mass of the Higgs boson (see Eq.~(A.26)
in~\cite{RGE-analysis-1}, and the discussion in~\cite{strumia}).
With $O(\alpha)$ accuracy, $\mu_{\mbox{\rm{IR}},m_H^2} \sim 4~\gv$
$O(\alpha \alpha_s)$-order corrections shifts this number to $11~\gv$.
The last ingredient of the Higgs potential is the Higgs self-coupling
$\lambda$.  The corresponding value of the IR scale we denote as
$\mu_{\mbox{\rm{IR}},\lambda}$. With $O(\alpha)+O(\alpha
\alpha_s)$ accuracy, $\mu_{\mbox{\rm{IR}},\lambda} \sim 164.55~\gv$,
and $\mu_{\mbox{\rm{IR}},\lambda} \sim 152.65~\gv$,
correspondingly. The inclusion the leading $O(\alpha^2)$ order
corrections from~\cite{RGE-analysis-2} stabilizes the
$\mu_{\mbox{\rm{IR}},\lambda}$ around $\sim 154~\gv$. At this scale,
\begin{equation}
\lambda(\mu_{\mbox{\rm{IR}},\lambda}) 
= \sqrt{2} G_F  3 M_H^2
= 3 \left( \frac{M_H}{246.22~\gv} \right)^2.
\label{L-IR}
\end{equation}
%
%
%
%
%

%

The IR-scale $\mu_{\mbox{\rm{IR}},EW}$ follows from the minimization
of the values of the matching conditions for the parameters of the Higgs
potential and lies in the interval $M_Z \leq \mu_{\mbox{\rm{IR}},EW}
\leq 2 M_W$.  At the boundary points of this interval, the quantity $
\delta \lambda(\mu^2) = \frac{\lambda(\mu^2)}{\Lambda_0}-1 \;,
$
with 
$
\Lambda_0 \equiv 3 \left( \frac{M_H}{246.22~\gv}\right)^2 \;, 
$
changes from 
$$
\left. 
\delta \lambda(\mu^2)
\right|_{\mu \sim M_Z}
\approx 
\frac{m_H^2(\mu_{\mbox{\rm{IR}},G_F}^2)}{M_H^2} - 1 
\Longrightarrow 
\left. 
\delta \lambda(\mu^2)
\right|_{\mu \sim 2 M_W}
\sim 0 \;.
$$ 
The existence of such an IR scale, numerically close to vector
boson masses, may be relevant for the extraction of
the Higgs self-coupling from cross-sections~\cite{higgs-selfcoupling}
as well as for construction of effective Lagrangian~\cite{effective}.

\noindent
\textbf{Acknowledgments:}
MYK is grateful to Sven Moch for formulation of the problem and for useful discussions;
to S.~Alekhin for clarifying of some aspects of \cite{ADM-analysis}; 
to F.~Bezrukov and  M.~Shaposhnikov for invitation to joint to project \cite{RGE-analysis-1}.
MYK would like to thank the organizers of the conference
{\it DIS2013} and specially to conveners
of ``Heavy Flavours'' Working Groups, S.~Naumann-Emme, M.~Sauter and  A.~Szczurek, 
for the invitation and for creating such a stimulating atmosphere.

\end{document}